\documentclass[eqsecnum,superscriptaddress,nofootinbib,11pt]{revtex4}
\usepackage{amsmath}
   \usepackage{bm}
\usepackage{amsthm,amscd}
\usepackage{mathrsfs}
\usepackage{verbatim}
\usepackage{amsfonts,amsmath,latexsym,amssymb,txfonts,bm}
\usepackage[american]{babel}
\usepackage{dsfont}
\usepackage[dvips]{graphicx}
\usepackage{latexsym}
\usepackage{epsfig}
\newcommand{\diff}{\mathrm{d}}

\def\beq{\begin{eqnarray}}
\def\eeq{\end{eqnarray}}

\begin{document}
\title{Expansion of Infinite Series Containing Modified Bessel Functions of the Second Kind}

\author{Guglielmo Fucci\footnote{Electronic address: fuccig@ecu.edu}}
\affiliation{Department of Mathematics, East Carolina University, Greenville, NC 27858 USA}

\author{Klaus Kirsten\footnote{Electronic address: Klaus\textunderscore Kirsten@Baylor.edu}}
\affiliation{GCAP-CASPER, Department of Mathematics, Baylor University, Waco, TX 76798 USA}

\date{\today}
\vspace{2cm}
\begin{abstract}

The aim of this work is to analyze general infinite sums containing modified Bessel functions of the second kind.
In particular we present a method for the construction of a proper asymptotic expansion for such series valid when one of the parameters in the
argument of the modified Bessel function of the second kind is small compared to the others. We apply the results obtained for the asymptotic expansion to
specific problems that arise in the ambit of quantum field theory.

\end{abstract}
\maketitle

\section{Introduction}

In quantum field theory the one-loop effective action is expressed in terms of the functional
determinant of the (elliptic and self-adjoint) operator of small disturbances \cite{bytse03,dewitt03}.
Since the real eigenvalues of an elliptic self-adjoint operator $L$ grow without bound, a na\"{i}ve computation of its
functional determinant would give, as a result, a meaningless infinite quantity. A regularized expression for the functional determinant of the operator $L$
can be obtained by using the spectral zeta function \cite{bytse03,elizalde94,hawking77,kirsten01,ray71}.
By denoting with $\lambda_{n}$, $n\in\mathbb{N}$, the real spectrum of $L$, which acts on suitable functions
defined on a smooth compact Riemannian manifold $M$, the spectral zeta function is defined as the following sum
\begin{equation}\label{0}
\zeta(s)=\sum_{n=1}^{\infty}\lambda_{n}^{-s}\;,
\end{equation}
where $s\in\mathbb{C}$. The above series is well defined in the half-plane $\Re(s)>D/2$, where $D$ represents the dimension of the manifold $M$,
and can be analytically continued in the entire complex plane to a meromorphic function possessing only simple poles \cite{mina49}. The zeta regularized functional
determinant of the operator $L$ is then expressed in terms of the derivative at $s=0$ of the associated (analytically continued) spectral zeta function.

Different types of methods have been developed with the purpose of finding the analytic continuation of the spectral zeta function (see e.g. \cite{elizalde95,kirsten01}).
The choice of which specific technique to use, however, depends heavily on whether or not the spectrum of the operator under consideration is explicitly
known. The analytically continued expression of a certain class of spectral zeta functions constructed from explicitly known spectra
often contains either single or double infinite series whose terms include modified Bessel functions of the second kind $K_{\nu}(z)$ \cite{elizalde94,elizalde95}.
The double Bessel series that appears in most applications has the general form
\begin{equation}\label{1}
f(s,\beta,B)=\sum_{n=1}^{\infty}\sum_{m=1}^{\infty}\left(\frac{m\beta}{\alpha_{n}}\right)^{s}\cos(2\pi m B)K_{-s}(2\alpha_{n}m\beta)\;,
\end{equation}
where $\beta>0$ is a dimensionless parameter, $B\in\mathbb{R}$ or $B\in i\mathbb{R}$, and the term $\alpha_{n}$ represents an increasing sequence of real numbers. The most commonly encountered Bessel series are
\begin{equation}\label{2}
h(s,\beta,B)=\sum_{m=1}^{\infty}(m\beta)^{s}\cos(2\pi m B)K_{-s}(2m\beta)\;,
\end{equation}
which can be obtained from (\ref{1}) by setting $\alpha_{n}=1$, and
\begin{equation}\label{3}
g(s,\beta)={\sum_{{\bf n}\in \mathbb{Z}^{d}}}^{\prime}\left(\frac{\beta}{|{\bf n}|}\right)^{s}K_{-s}(2|{\bf n}|\beta)\;,
\end{equation}
where $|{\bf n}|=(n_{1}^{2}+\cdots+n_{d}^{2})^{1/2}$ and the prime indicates the omission of the origin of $\mathbb{Z}^{d}$. We would like to point out that infinite series of the form displayed above do not arise only in the analytically continued expressions of spectral zeta functions but they can also be found, for instance,
in problems leading to lattice sums \cite{singh89}.

From the expressions (\ref{1})-(\ref{3}) it is not very difficult to realize that the series converge quickly
when $\beta\gg1$ by virtue of the following asymptotic behavior of the modified Bessel function of the second kind
\begin{equation}\label{4}
K_{\nu}(z)\sim\sqrt{\frac{\pi}{2z}}e^{-z}\;,
\end{equation}
for $z\to\infty$ with $\nu$ fixed \cite{gradshtein07}. This implies that $f(s,\beta,B)$, $h(s,\beta,B)$, and $g(s,\beta)$ become very useful for numerical evaluations when
$\beta$ is large. For $\beta\ll1$, instead, the expressions (\ref{1})-(\ref{3}) lose their accuracy.
This is quite undesirable since it is often of particular interest to obtain an asymptotic expansion of the series (\ref{1})-(\ref{3})
valid when the parameter $\beta$ is small . For instance, in the study of finite temperature effects in quantum field theory on manifolds,
the parameter $\beta$ is proportional to the inverse of the temperature $T$ \cite{campore90,dowker84}. Therefore, the small $\beta$ expansion describes
the high temperature behavior of the system under consideration \cite{elizalde95}. In addition, series of the form (\ref{1})
are obtained when studying the zeta regularized one-loop effective action of fields propagating on Kaluza-Klein manifolds of the type $M\times S^{1}$ \cite{elizalde94}.
In this case the parameter $\beta$ is related to the radius of the compactified dimension.

Due to its importance, the small-$\beta$ expansion of series of the form (\ref{1})-(\ref{3}) has been analyzed by several authors.
Unfortunately, however, this expansion has not always been performed correctly.
In fact, in many occasions the small-$\beta$ expansion of (\ref{1})-(\ref{3}) has been obtained first by substituting
the small argument expansion of the modified
Bessel functions of the second kind in the series (\ref{1})-(\ref{3}) and then by erroneously
regularizing the ensuing divergent expression with a suitable zeta function. The problem with this approach
lies in the fact that the argument of the modified Bessel functions of the second kind that appears in the series (\ref{1})-(\ref{3})
{\it cannot} be assumed to be small for all $n$. In fact, from the expressions (\ref{1})-(\ref{3}) it is easy to realize that the argument of the
modified Bessel functions of the second kind has the general form $a_{n}\beta$ where $a_{n}$, $n\in\mathbb{N}^{+}$, is an increasing
sequence of positive real numbers. Now, for any small $\beta$ there always exists an integer $m\in\mathbb{N}^{+}$ such that $a_{n}\beta>1$
for any $n> m$. This implies that the argument of the modified Bessel function fails to be small {\it uniformly} in $n$.
This fact seems to have been overlooked also in the study of problems leading to spectral zeta functions
of the form
\begin{equation}\label{5}
\sum_{n=0}^{\infty}\sum_{m=1}^{\infty}\left[(n+c)^{2}+a_{n}^{2}\beta\right]^{-s}\;,
\end{equation}
where $c\in\mathbb{R}^{+}$ and $a_{n}$ denotes, for instance, the eigenvalues of an elliptic self-adjoint
differential operator. In this framework the small-$\beta$ expansion of (\ref{5}) has been obtained
by erroneously assuming that the product $a_{n}^{2}\beta$ is small for all values of $n$. It is important to mention, at this point, that even though
the small-$\beta$ expansion of series of the form (\ref{1})-(\ref{3}) has been often a source of confusion, examples of properly performed expansions
can be found, for instance, in 
\cite{acto87-35-793,weld86-270-79,allen86,elizalde94,kirsten92}. 

The aim of this work is to present a method for obtaining the correct small-$\beta$ expansion of series of
the form (\ref{1})-(\ref{3}) which is valid for arbitrary values of $B$ and $s$. Although, as mentioned above,
correct small-$\beta$ expansions of (\ref{1})-(\ref{3}) have already been obtained, they are scattered in the literature and, most importantly,
they have been found mostly for special cases. With this work we want to provide a quite general method for the small-$\beta$ expansion
which can be used to find expansions performed for special cases and can serve as a useful tool for future research in this area.

The outline of the paper is as follows. In the next two sections we utilize a complex integral representation
of the modified Bessel function of the second kind to obtain, through Cauchy's residue theorem, the small-$\beta$
expansion of the series (\ref{1})-(\ref{3}). In section \ref{sec3} we apply our general results to special
cases that are of wide interest in physical applications. In the last section we summarize and discuss
the main results of this work.

\section{Small Parameter Expansion of the Double Bessel Series}\label{sec2}

In this section we develop the small-$\beta$ expansion of the double Bessel series (\ref{1}).
Due to the exponential decay of the modified Bessel function of the second kind, the double-series (\ref{1})
converges for all values of $s$ and, hence, defines an entire function in the finite complex  $s$-plane. Clearly,
the same remark also applies to the series (\ref{2}) and (\ref{3})
which allows us to conclude that also (\ref{2}) and (\ref{3}) are entire functions in the finite complex $s$-plane. In order to obtain an expression for (\ref{1})
that is suitable for a small-$\beta$ expansion we use the following complex integral representation
\begin{equation}\label{6}
K_{s}(z)=\left(\frac{z}{2}\right)^{s}\frac{1}{4\pi i}\int_{c-i\infty}^{c+i\infty}\Gamma(t)\Gamma(t-s)\left(\frac{z}{2}\right)^{-2t}\diff t\;,
\end{equation}
with $c>\textrm{max}\{0,\Re(s)\}$ \cite{gradshtein07}, in the function $f(s,\beta,B)$ to obtain
\begin{equation}\label{7}
f(s,\beta,B)=\frac{1}{4\pi i}\sum_{n=1}^{\infty}\sum_{m=1}^{\infty}\cos(2\pi m B)\int_{c-i\infty}^{c+i\infty}\Gamma(t)\Gamma(t+s)(\beta m)^{-2t}\alpha_{n}^{-2t-2s}\diff t\;.
\end{equation}

By assuming that the increasing sequence of positive real numbers $\{\alpha_{n}\}_{n\in\mathbb{N}^{+}}$ represents the eigenvalues
of a Laplace-type operator on a manifold $M$ of dimension $D$,
we define the associated spectral zeta function as follows
\begin{equation}\label{8}
\zeta_{M}(s)=\sum_{n=1}^{\infty}\alpha_{n}^{-s}\;,
\end{equation}
which is valid for $\Re(s)>D/2$. The function (\ref{8}) can be analytically continued to a meromorphic function with simple poles at
$s=(D-k)/2$ with $k=\{0,\ldots,D-1\}$ and $s=-(2l+1)/2$ with $l\in\mathbb{N}_{0}$ \cite{seel68-10-288}.
In addition, for $\Re(s)>1/2$ it is not very difficult to obtain
\begin{equation}\label{9}
\sum_{m=1}^{\infty}m^{-2s}\cos(2\pi m B)=\frac{1}{2}\left[\textrm{Li}_{2s}\left(e^{2\pi i B}\right)+\textrm{Li}_{2s}\left(e^{-2\pi i B}\right)\right]\;,
\end{equation}
where $\textrm{Li}_{\nu}(z)$ represents the polylogarithimc function. For $\nu<1$
the polylogarithm has a branch point at $z=1$ \cite{erdelyi53}. Therefore, it is convenient to distinguish between the case $B\neq 0$ and $B=0$.
For $\Re(s)>1/2$ and $B\neq 0$ we use the relation (\ref{9}), for $\Re(s)>1/2$ and $B=0$ the series on the left-hand-side of (\ref{9}) becomes instead the Riemann zeta function
$\zeta_{R}(2s)$.

For $c>\textrm{max}\{1/2,D/2-\Re(s)\}$  we can interchange the sum and the integral in (\ref{7}) and
then use the definition (\ref{8}) and the result (\ref{9}) to write, for $B\neq 0$,
\begin{equation}\label{10}
f(s,\beta,B)=\frac{1}{8\pi i}\int_{c-i\infty}^{c+i\infty}\Gamma(t)\Gamma(t+s)\beta^{-2t}\zeta_{M}(t+s)
\left[\textrm{Li}_{2t}\left(e^{2\pi i B}\right)+\textrm{Li}_{2t}\left(e^{-2\pi i B}\right)\right]\diff t\;.
\end{equation}
For $B=0$ we have, instead, the following expression
\begin{equation}\label{11}
f(s,\beta,0)=\frac{1}{4\pi i}\int_{c-i\infty}^{c+i\infty}\Gamma(t)\Gamma(t+s)\beta^{-2t}\zeta_{M}(t+s)\zeta_{R}(2t)\diff t\;.
\end{equation}
The integral representations (\ref{10}) and (\ref{11}) are particularly suitable for performing an expansion as $\beta\to 0$.
This expansion is obtained by closing the integration contour to the left and by computing the resulting integral by using Cauchy's residue theorem.
For the case $B\neq 0$ the terms in the integrand of (\ref{10}) have poles at the points
\begin{equation}\label{12}
t=-s+\frac{D-k}{2}\;,\quad t=-s-\frac{2l+1}{2}\;,\quad t=-s-l\;,\quad t=-l\;,
\end{equation}
with $k=\{0,\ldots,D-1\}$ and $l\in\mathbb{N}_{0}$. From (\ref{12}) it is clear that depending on the value of $s$, the integrand of (\ref{10})
might have either simple or double poles. In fact, it is not difficult to realize that for $s\neq k$, with $k\in\mathbb{Z}$, and for
$s\neq\pm (2n+1)/2$, $n\in\mathbb{N}_{0}$, none of the points in (\ref{12}) coincide and, therefore,
the integrand in (\ref{10}) possesses only simple poles. When either $s=k$ or $s=\pm (2n+1)/2$ the integrand in (\ref{10})
will develop, in addition to simple poles, also double poles.

For $s\neq k$ and $s\neq\pm (2n+1)/2$, we close the contour of integration in (\ref{10}) to the left as to include the simple poles of the integrand. From the poles at $t=-s+(D-k)/2$ we obtain the following contribution
\begin{equation}\label{13}
f_{1}(s,\beta,B)=\frac{\beta^{2s}}{4}\sum_{k=0}^{D-1}\Gamma\left(-s+\frac{D-k}{2}\right)\Gamma\left(\frac{D-k}{2}\right)\beta^{k-D}
\textrm{Res}\,\zeta_{M}\left(\frac{D-k}{2}\right)\left[\textrm{Li}_{-2s+D-k}\left(e^{2\pi i B}\right)+\textrm{Li}_{-2s+D-k}\left(e^{-2\pi i B}\right)\right]\;.
\end{equation}
Since the residues of the spectral zeta function are proportional to the coefficients $A^{M}_{n/2}$ of the asymptotic expansion of the trace of the
heat kernel associated with the Laplace-type operator on $M$ as follows \cite{seel68-10-288}
\begin{equation}\label{14}
A^{M}_{\frac{k}{2}}=\Gamma\left(\frac{D-k}{2}\right)\textrm{Res}\,\zeta_{M}\left(\frac{D-k}{2}\right)\;,
\end{equation}
and
\begin{equation}\label{15}
A^{M}_{\frac{D+2l+1}{2}}=\Gamma\left(-\frac{2l+1}{2}\right)\textrm{Res}\,\zeta_{M}\left(-\frac{2l+1}{2}\right)\;,
\end{equation}
the expression in (\ref{13}) becomes
\begin{equation}\label{16}
f_{1}(s,\beta,B)=\frac{\beta^{2s}}{4}\sum_{k=0}^{D-1}\Gamma\left(-s+\frac{D-k}{2}\right)\beta^{k-D}A^{M}_{\frac{k}{2}}\left[\textrm{Li}_{-2s+D-k}\left(e^{2\pi i B}\right)+\textrm{Li}_{-2s+D-k}\left(e^{-2\pi i B}\right)\right]\;.
\end{equation}

From the poles at $t=-s-(2l+1)/(2)$ we have the following contribution
\begin{equation}\label{17}
f_{2}(s,\beta,B)=\frac{\beta^{2s+1}}{4}\sum_{l=0}^{\infty}\Gamma\left(-s-\frac{2l+1}{2}\right)\beta^{2l}A^{M}_{\frac{D+2l+1}{2}}\left[\textrm{Li}_{-2s-2l-1}\left(e^{2\pi i B}\right)+\textrm{Li}_{-2s-2l-1}\left(e^{-2\pi i B}\right)\right]\;,
\end{equation}
which has been obtained by using the relation (\ref{15}).

The residues associated with the poles located at $t=-s-l$ contribute
\begin{equation}\label{18}
f_{3}(s,\beta,B)=\frac{\beta^{2s}}{4}\sum_{l=0}^{\infty}\frac{(-1)^{l}}{l!}\Gamma(-s-l)\beta^{2l}\zeta_{M}(-l)
\left[\textrm{Li}_{-2s-2l}\left(e^{2\pi i B}\right)+\textrm{Li}_{-2s-2l}\left(e^{-2\pi i B}\right)\right]\;.
\end{equation}
By using the relation \cite{seel68-10-288}
\begin{equation}\label{19}
\zeta_{M}(-l)=(-1)^{l}l! A_{\frac{D}{2}+l}^{M}\;,
\end{equation}
in (\ref{18}) we obtain
\begin{equation}\label{20}
f_{3}(s,\beta,B)=\frac{\beta^{2s}}{4}\sum_{l=0}^{\infty}\Gamma(-s-l)\beta^{2l}A_{\frac{D}{2}+l}^{M}
\left[\textrm{Li}_{-2s-2l}\left(e^{2\pi i B}\right)+\textrm{Li}_{-2s-2l}\left(e^{-2\pi i B}\right)\right]\;.
\end{equation}

Finally, from the poles at $t=-l$ we have
\begin{equation}\label{21}
f_{4}(s,\beta,B)=\frac{1}{4}\sum_{l=0}^{\infty}\frac{(-1)^{l}}{l!}\Gamma(s-l)\beta^{2l}\zeta_{M}(s-l)
\left[\textrm{Li}_{-2l}\left(e^{2\pi i B}\right)+\textrm{Li}_{-2l}\left(e^{-2\pi i B}\right)\right]\;.
\end{equation}
By noticing that for $n\in\mathbb{N}_{0}$ the polylogarithmic function satisfies the following relations \cite{erdelyi53}
\begin{equation}\label{22}
\textrm{Li}_{n}\left(e^{2\pi i x}\right)+\textrm{Li}_{n}\left(e^{-2\pi i x}\right)=-\frac{(2\pi i)^{n}}{n!}B_{n}(x)\;,
\end{equation}
with $B_{n}(x)$ denoting the Bernoulli polynomials, and for $m\in\mathbb{N^{+}}$
\begin{equation}\label{23}
\textrm{Li}_{-m}\left(e^{2\pi i B}\right)=(-1)^{m+1}\textrm{Li}_{-m}\left(e^{-2\pi i B}\right)\;,
\end{equation}
we can conclude that the sum of polylogarithmic functions in (\ref{21}) vanishes identically except when $l=0$. The last remarks
allow us to write that
\begin{equation}\label{24}
f_{4}(s,\beta,B)=-\frac{1}{4}\Gamma(s)\zeta_{M}(s)\;.
\end{equation}

At this point, by adding the results obtained in (\ref{16}), (\ref{17}), (\ref{20}), and (\ref{24}) we have
\begin{eqnarray}\label{25}
f(s,\beta,B)\sim\frac{\beta^{2s}}{4}\sum_{l=-D}^{\infty}\Gamma\left(-s-\frac{l}{2}\right)\beta^{l}A_{\frac{D+l}{2}}^{M}
\left[\textrm{Li}_{-2s-l}\left(e^{2\pi i B}\right)+\textrm{Li}_{-2s-l}\left(e^{-2\pi i B}\right)\right]
-\frac{1}{4}\Gamma(s)\zeta_{M}(s)\;,
\end{eqnarray}
where, here and in the rest of this paper, we use $\sim$ to represent the small-$\beta$ asymptotic expansion.

When either $s=k$, $k\in\mathbb{Z}$, or $s=\pm (2n+1)/2$, $n\in\mathbb{N}_{0}$, the integrand in (\ref{10}) will develop
both simple and double poles. The contributions coming from the double poles can be easily computed by using the following argument.
The integrand appearing in the integral representations utilized in this work can be viewed as a product of three functions, $I(t)=p(t)q(t)r(t)$.
One of the functions, which we assume to be $r(t)$, is analytic while $p(t)$ and $q(t)$ possess $N$, finitely or infinitely many, coinciding simple poles at $t=c_{i}$.
The residue of each resulting double pole at $t=c_{i}$ can then be computed to be
\begin{equation}\label{25a}
\textrm{Res}\,I(c_{i})=(\textrm{Res}\,p(c_{i}))(\textrm{FP}\,q(c_{i})) r(c_{i})+(\textrm{Res}\,q(c_{i}))(\textrm{FP}\,p(c_{i})) r(c_{i})+(\textrm{Res}\,q(c_{i}))(\textrm{Res}\,p(c_{i}))r^{\prime}(c_{i})\;.
\end{equation}

We start by analyzing the case $s=n$, with $n\in\mathbb{N}^{+}$, and an even-dimensional manifold $M$, namely $D=2d$. Under this assumption the integrand in (\ref{10}) has
simple poles for $t=-n-(2m+1)/2$, $m\in\mathbb{N}_{0}$, and for $t=-n+d-p-1/2$, $p=\{0,\ldots,d-1\}$, and double poles for $t=-q$, where $q\in\mathbb{N}^{+}$ such that $q\geq n$.
Moreover, if $0<n\leq d$, $t=-n+d-j$ represents a simple pole for $j=\{0,\ldots,d-n-1\}$ and a double pole when $j=\{d-n,\ldots,d-1\}$.
If $n\geq d+1$, $t=-n+d-j$ is a double pole for $j=\{0,\ldots,d-1\}$, and $t=-q$ with $q=\{0,\ldots,n-d-1\}$, represents instead a simple pole.
The integral in (\ref{10}) is computed by closing the integration contour to the left. The contributions coming from the simple poles can be computed as before
while the contributions from the double poles are found by using the argument leading to (\ref{25a}). This gives the result
\begin{eqnarray}\label{25b}
\lefteqn{f_{D=2d}(n,\beta,B)\sim\frac{1}{4}\left[-(n-1)!\textrm{FP}\,\zeta_{M}(n)+A_{d-n}^{M}(\gamma-\Psi(n)+2\ln\beta)\right]\Theta(d-n)}\nonumber\\
& &+\Theta(d-n-1)\frac{\beta^{2n-2d}}{4}\sum_{j=0}^{d-n-1}\Gamma(-n+d-j)\beta^{2j}A_{j}^{M}\left[\textrm{Li}_{-2n+2d-2j}\left(e^{2\pi i B}\right)+\textrm{Li}_{-2n+2d-2j}\left(e^{-2\pi i B}\right)\right]\nonumber\\
&&+\frac{\beta^{2n+1}}{4}\sum_{m=-d}^{\infty}\Gamma\left(-n-\frac{2m+1}{2}\right)\beta^{2m}A^{M}_{d+m+\frac{1}{2}}\left[\textrm{Li}_{-2n-2m-1}\left(e^{2\pi i B}\right)+\textrm{Li}_{-2n-2m-1}\left(e^{-2\pi i B}\right)\right]\nonumber\\
&&+\frac{\beta^{2n-2d}}{2}\sum_{j=\max\{d-n,0\}}^{d-1}\frac{(-1)^{n-d+j}}{(n-d+j)!}\beta^{2j}A^{M}_{j}\left[\textrm{Li}^{\prime}_{-2n+2d-2j}\left(e^{2\pi i B}\right)+\textrm{Li}^{\prime}_{-2n+2d-2j}\left(e^{-2\pi i B}\right)\right]\nonumber\\
&&+\frac{1}{2}\sum_{m=n}^{\infty}\frac{(-1)^{m}}{m!}\beta^{2m}A^{M}_{d+m-n}\left[\textrm{Li}^{\prime}_{-2m}\left(e^{2\pi i B}\right)+\textrm{Li}^{\prime}_{-2m}\left(e^{-2\pi i B}\right)\right]-\frac{1}{4}(n-1)!\zeta_{M}(n)\Theta(n-d-1)\;,
\end{eqnarray}
where $\Theta(x)$ is the unit step-function, $\gamma$ is the Euler-Mascheroni constant, $\Psi(x)$ represents the logarithmic derivative of the gamma function,
and $\textrm{Li}^{\prime}_{s}(z)=\partial_{s}\textrm{Li}_{s}(z)$. For $s=n$, with $n\in\mathbb{N}^{+}$, and $D=2d+1$, the integrand in (\ref{10})
presents simple poles for $t=-n-(2m+1)/2$, $m\in\mathbb{N}_{0}$, and for $t=-n+d-j-1/2$, $j=\{0,\ldots,d\}$, and double poles for $t=-q$, where $q\in\mathbb{N}^{+}$ with $q\geq n$. In addition, if $0<n\leq d$, $t=-n+d-p$ is a simple pole for $p=\{0,\ldots,d-n\}$ and a double pole when $j=\{d-n,\ldots,d-1\}$. If $n\geq d+1$, $t=-n+d-p$ is a double pole for $j=\{0,\ldots,d-1\}$ and $t=-q$, with $q=\{0,\ldots,n-d-1\}$, is a simple pole. The application of Cauchy's residue theorem then gives
\begin{eqnarray}\label{25c}
\lefteqn{f_{D=2d+1}(n,\beta,B)\sim\frac{1}{4}\left[-(n-1)!\textrm{FP}\,\zeta_{M}(n)+A_{d-n+\frac{1}{2}}^{M}(\gamma-\Psi(n)+2\ln\beta)\right]\Theta(d-n)}\nonumber\\
&&+\Theta(d-n-1)\frac{\beta^{2n-2d}}{4}\sum_{p=0}^{d-n-1}\Gamma(-n+d-p)\beta^{2p}A_{p+\frac{1}{2}}^{M}\left[\textrm{Li}_{-2n+2d-2p}\left(e^{2\pi i B}\right)+\textrm{Li}_{-2n+2d-2p}\left(e^{-2\pi i B}\right)\right]\nonumber\\
&&+\frac{\beta^{2n+1}}{4}\sum_{m=-d-1}^{\infty}\Gamma\left(-n-\frac{2m+1}{2}\right)\beta^{2m}A^{M}_{d+m+1}\left[\textrm{Li}_{-2n-2m-1}\left(e^{2\pi i B}\right)+\textrm{Li}_{-2n-2m-1}\left(e^{-2\pi i B}\right)\right]\nonumber\\
&&+\frac{\beta^{2n-2d}}{2}\sum_{p=\max\{d-n,0\}}^{d-1}\frac{(-1)^{n-d+p}}{(n-d+p)!}\beta^{2p}A^{M}_{p+\frac{1}{2}}\left[\textrm{Li}^{\prime}_{-2n+2d-2p}\left(e^{2\pi i B}\right)+\textrm{Li}^{\prime}_{-2n+2d-2p}\left(e^{-2\pi i B}\right)\right]\nonumber\\
&&+\frac{1}{2}\sum_{m=n}^{\infty}\frac{(-1)^{m}}{m!}\beta^{2m}A^{M}_{d+m-n+\frac{1}{2}}\left[\textrm{Li}^{\prime}_{-2m}\left(e^{2\pi i B}\right)+\textrm{Li}^{\prime}_{-2m}\left(e^{-2\pi i B}\right)\right]-\frac{1}{4}(n-1)!\zeta_{M}(n)\Theta(n-d-1)\;.
\end{eqnarray}
When $s=-n$ with $n\in\mathbb{N}_{0}$ for both even and odd-dimensional manifolds $M$, the integrand in (\ref{10}) contains
simple poles at $t=n+(D-k)/2$, with $k=\{0\ldots,D-1\}$, at $t=n-(2m+1)/2$, $m\in\mathbb{N}_{0}$, and at $t=q$ with $q=\{1,\ldots, n\}$.
Double poles appear, instead, at the points $t=-q$ with $q\in\mathbb{N}_{0}$. Taking into account the above poles, the integral in (\ref{10}) leads to
the result
\begin{eqnarray}\label{25d}
\lefteqn{f(-n,\beta,B)\sim\frac{1}{4}A^{M}_{\frac{D}{2}+n}\left[\gamma-\Psi(1+n)+2\ln\beta\right]-\frac{1}{4}\frac{(-1)^{n}}{n!}\zeta_{M}^{\prime}(-n)}\nonumber\\
&&+\frac{\beta^{-2n-D}}{4}\sum_{k=0}^{D-1}\Gamma\left(n+\frac{D-k}{2}\right)\beta^{k}A^{M}_{\frac{k}{2}}\left[\textrm{Li}_{2n+D-k}\left(e^{2\pi i B}\right)+\textrm{Li}_{2n+D-k}\left(e^{-2\pi i B}\right)\right]\nonumber\\
&&+\frac{\beta^{-2n+1}}{4}\sum_{m=0}^{\infty}\Gamma\left(n-\frac{2m+1}{2}\right)\beta^{2m}A^{M}_{\frac{D+2m+1}{2}}\left[\textrm{Li}_{2n-2m-1}\left(e^{2\pi i B}\right)+\textrm{Li}_{2n-2m-1}\left(e^{-2\pi i B}\right)\right]\nonumber\\
&&+\frac{1}{2}\sum_{m=0}^{\infty}\frac{(-1)^{m}}{m!}\beta^{2m}A^{M}_{\frac{D}{2}+m+n}\left[\textrm{Li}^{\prime}_{-2m}\left(e^{2\pi i B}\right)+\textrm{Li}^{\prime}_{-2m}\left(e^{-2\pi i B}\right)\right]\nonumber\\
&&+\frac{\Theta(n-1)}{4}\sum_{m=1}^{n}(m-1)!\beta^{-2m}A^{M}_{\frac{D}{2}+n-m}\left[\textrm{Li}_{2m}\left(e^{2\pi i B}\right)+\textrm{Li}_{2m}\left(e^{-2\pi i B}\right)\right]\;.
\end{eqnarray}

We consider next the values $s=(2n+1)/2$, $n\in\mathbb{N}_{0}$, and $D=2d$. In this case the integrand in (\ref{10}) has simple poles
at $t=-n-(2m+1)/2$, with $m\in\mathbb{N}_{0}$, at $t=-n+d-j-1/2$, with $j=\{0,\ldots,d-1\}$, and double poles at $t=-q$, where $q\in\mathbb{N}^{+}$ with $q\geq n+1$.
In addition, if $0\leq n\leq d-1$, then $t=-n+d-p-1$ is a simple pole for $p=\{0,\ldots,d-n-2\}$ and a double pole for $p=\{d-n-1,\ldots,d-1\}$.
If instead $n\geq d$, then $t=-n+d-p-1$ represents a double pole for $p=\{0,\ldots,d-1\}$ and $t=-q$,
with $q=\{0,\ldots,n-d\}$, is a simple pole. By computing (\ref{10}) in this case we obtain
\begin{eqnarray}\label{25e}
\lefteqn{f_{D=2d}\left(\frac{2n+1}{2},\beta,B\right)\sim\frac{1}{4}\left[-\Gamma\left(n+\frac{1}{2}\right)\textrm{FP}\,\zeta_{M}\left(n+\frac{1}{2}\right)+A_{d-n-\frac{1}{2}}^{M}\left(\gamma-\Psi\left(n+\frac{1}{2}\right)+2\ln\beta\right)\right]\Theta(d-n-1)}\nonumber\\
&&+\Theta(d-n-2)\frac{\beta^{2n-2d+2}}{4}\sum_{j=0}^{d-n-2}\Gamma(-n+d-j-1)\beta^{2j}A_{j+\frac{1}{2}}^{M}\left[\textrm{Li}_{-2n+2d-2j-2}\left(e^{2\pi i B}\right)+\textrm{Li}_{-2n+2d-2j-2}\left(e^{-2\pi i B}\right)\right]\nonumber\\
&&+\frac{\beta^{2n+1}}{4}\sum_{m=-d}^{\infty}\Gamma\left(-n-\frac{2m+1}{2}\right)\beta^{2m}A^{M}_{d+m}\left[\textrm{Li}_{-2n-2m-1}\left(e^{2\pi i B}\right)+\textrm{Li}_{-2n-2m-1}\left(e^{-2\pi i B}\right)\right]\nonumber\\
&&+\frac{\beta^{2n-2d+2}}{2}\sum_{j=\max\{d-n-1,0\}}^{d-1}\frac{(-1)^{n-d+j+1}}{(n-d+j+1)!}\beta^{2j}A^{M}_{j+\frac{1}{2}}\left[\textrm{Li}^{\prime}_{-2n+2d-2j-2}\left(e^{2\pi i B}\right)+\textrm{Li}^{\prime}_{-2n+2d-2j-2}\left(e^{-2\pi i B}\right)\right]\nonumber\\
&&+\frac{1}{2}\sum_{m=n+1}^{\infty}\frac{(-1)^{m}}{m!}\beta^{2m}A^{M}_{d+m-n-\frac{1}{2}}\left[\textrm{Li}^{\prime}_{-2m}\left(e^{2\pi i B}\right)+\textrm{Li}^{\prime}_{-2m}\left(e^{-2\pi i B}\right)\right]-\frac{1}{4}\Gamma\left(n+\frac{1}{2}\right)\zeta_{M}\left(n+\frac{1}{2}\right)\Theta(n-d)\;,
\end{eqnarray}
For $s=(2n+1)/2$, $n\in\mathbb{N}_{0}$, and $D=2d+1$ one finds simple poles at $t=-n-(2m+1)/2$, with $m\in\mathbb{N}_{0}$,
at $t=-n+d-j-1/2$, with $j=\{0,\ldots,d-1\}$, and double poles at $t=-q$, where $q\in\mathbb{N}^{+}$ with $q\geq n+1$.
Furthermore, if $0\leq n\leq d$, then $t=-n+d-j$ is a simple pole for $j=\{0,\ldots,d-n-1\}$ and a double pole for $j=\{d-n,\ldots,d\}$.
If $n\geq d+1$, then $t=-n+d-j$ is a double pole for $j=\{0,\ldots,d-1\}$ and $t=-q$, with $q=\{0,\ldots,n-d-1\}$, is a simple pole.
In this case the integral in (\ref{10}) gives
\begin{eqnarray}\label{25f}
\lefteqn{f_{D=2d+1}\left(\frac{2n+1}{2},\beta,B\right)\sim\frac{1}{4}\left[-\Gamma\left(n+\frac{1}{2}\right)\textrm{FP}\,\zeta_{M}\left(n+\frac{1}{2}\right)
+A_{d-n}^{M}\left(\gamma-\Psi\left(n+\frac{1}{2}\right)+2\ln\beta\right)\right]\Theta(d-n)}\nonumber\\
&&+\Theta(d-n-1)\frac{\beta^{2n-2d}}{4}\sum_{j=0}^{d-n-1}\Gamma(-n+d-j)\beta^{2j}A_{j}^{M}\left[\textrm{Li}_{-2n+2d-2j}\left(e^{2\pi i B}\right)+\textrm{Li}_{-2n+2d-2j}\left(e^{-2\pi i B}\right)\right]\nonumber\\
&&+\frac{\beta^{2n+1}}{4}\sum_{m=-d}^{\infty}\Gamma\left(-n-\frac{2m+1}{2}\right)\beta^{2m}A^{M}_{d+m+\frac{1}{2}}\left[\textrm{Li}_{-2n-2m-1}\left(e^{2\pi i B}\right)+\textrm{Li}_{-2n-2m-1}\left(e^{-2\pi i B}\right)\right]\nonumber\\
&&+\frac{\beta^{2n-2d}}{2}\sum_{j=\max\{d-n,0\}}^{d}\frac{(-1)^{n-d+j}}{(n-d+j)!}\beta^{2j}A^{M}_{j}\left[\textrm{Li}^{\prime}_{-2n+2d-2j}\left(e^{2\pi i B}\right)+\textrm{Li}^{\prime}_{-2n+2d-2j}\left(e^{-2\pi i B}\right)\right]\nonumber\\
&&+\frac{1}{2}\sum_{m=n+1}^{\infty}\frac{(-1)^{m}}{m!}\beta^{2m}A^{M}_{d+m-n}\left[\textrm{Li}^{\prime}_{-2m}\left(e^{2\pi i B}\right)+\textrm{Li}^{\prime}_{-2m}\left(e^{-2\pi i B}\right)\right]-\frac{1}{4}\Gamma\left(n+\frac{1}{2}\right)\zeta_{M}\left(n+\frac{1}{2}\right)\Theta(n-d-1)\;,\nonumber\\
\end{eqnarray}
Lastly, for $s=-(2n+1)/2$, $n\in\mathbb{N}_{0}$, and for both even and odd-dimensional manifolds $M$, the integrand in (\ref{10})
has simple poles at $t=n+(D-k+1)/2$, with $k=\{0\ldots,D-1\}$, at $t=n-m+1/2$, for $m\in\mathbb{N}_{0}$, and at $t=-q$, with $q=\{1,\ldots,n\}$.
Double poles appear, instead, for $t=-q$ with $q\in\mathbb{N}_{0}$. In this case (\ref{10}) becomes
\begin{eqnarray}\label{25g}
\lefteqn{f\left(-\frac{2n+1}{2},\beta,B\right)\sim-\frac{1}{4}A^{M}_{\frac{D+2n+1}{2}}\left[\Psi\left(-\frac{2n+1}{2}\right)-\gamma-2\ln\beta\right]
-\frac{1}{4}\Gamma\left(-\frac{2n+1}{2}\right)\textrm{FP}\,\zeta_{M}\left(-\frac{2n+1}{2}\right)}\nonumber\\
&&+\frac{\beta^{-2n-D-1}}{4}\sum_{k=0}^{D-1}\Gamma\left(n+\frac{D-k+1}{2}\right)\beta^{k}A^{M}_{\frac{k}{2}}\left[\textrm{Li}_{2n+D-k+1}\left(e^{2\pi i B}\right)+\textrm{Li}_{2n+D-k+1}\left(e^{-2\pi i B}\right)\right]\nonumber\\
&&+\frac{\beta^{-2n-1}}{4}\sum_{m=0}^{\infty}\Gamma\left(n-m+\frac{1}{2}\right)\beta^{2m}A^{M}_{\frac{D}{2}+m}\left[\textrm{Li}_{2n-2m+1}\left(e^{2\pi i B}\right)+\textrm{Li}_{2n-2m+1}\left(e^{-2\pi i B}\right)\right]\nonumber\\
&&+\frac{1}{2}\sum_{m=0}^{\infty}\frac{(-1)^{m}}{m!}\beta^{2m}A^{M}_{\frac{D+1}{2}+m+n}\left[\textrm{Li}^{\prime}_{-2m}\left(e^{2\pi i B}\right)+\textrm{Li}^{\prime}_{-2m}\left(e^{-2\pi i B}\right)\right]\nonumber\\
&&+\frac{\Theta(n-1)}{4}\sum_{m=1}^{n}(m-1)!\beta^{-2m}A^{M}_{\frac{D+1}{2}+n-m}\left[\textrm{Li}_{2m}\left(e^{2\pi i B}\right)+\textrm{Li}_{2m}\left(e^{-2\pi i B}\right)\right]\;.
\end{eqnarray}

Let us now assume that $B=0$. In this case the relevant integral is $f(s,\beta,0)$ in (\ref{11}).
Once again, when  $s\neq k$, with $k\in\mathbb{Z}$, and when
$s\neq\pm (2n+1)/2$, $n\in\mathbb{N}_{0}$ it is not difficult to verify that the integrand in (\ref{11})
contains only simple poles which are located at the points given in (\ref{12}) and at the point $t=1/2$.
The integral $f(s,\beta,0)$ is then evaluated by closing the contour to the left and by using Cauchy's residue theorem.
From the poles positioned at $t=-s+(D-k)/2$ we obtain the following contribution
\begin{equation}\label{30}
f_{1}(s,\beta,0)=\frac{\beta^{2s}}{2}\sum_{k=0}^{D-1}\Gamma\left(-s+\frac{D-k}{2}\right)\beta^{k-D}A^{M}_{\frac{k}{2}}\zeta_{R}(-2s+D-k)\;.
\end{equation}
From the poles at $t=-s-(2l+1)/2$ we get
\begin{equation}\label{31}
f_{2}(s,\beta,0)=\frac{\beta^{2s+1}}{2}\sum_{l=0}^{\infty}\Gamma\left(-s-\frac{2l+1}{2}\right)\beta^{2l}A^{M}_{\frac{D+2l+1}{2}}\zeta_{R}(-2s-2l-1)\;,
\end{equation}
while from the ones at $t=-s-l$ we have
\begin{equation}\label{32}
f_{3}(s,\beta,0)=\frac{\beta^{2s}}{2}\sum_{l=0}^{\infty}\Gamma(-s-l)\beta^{2l}A_{\frac{D}{2}+l}^{M}
\zeta_{R}(-2s-2l)\;.
\end{equation}
Since $\zeta_{R}(-2k)=0$ for $k\in\mathbb{N}^{+}$, the only contribution coming from the poles at $t=-l$
can be found to be
\begin{equation}\label{33}
f_{4}(s,\beta,0)=-\frac{1}{4}\Gamma(s)\zeta_{M}(s)\;.
\end{equation}
Lastly, from the pole at $t=1/2$ we obtain
\begin{equation}\label{34}
f_{5}(s,\beta,0)=\frac{\sqrt{\pi}}{4\beta}\Gamma\left(s+\frac{1}{2}\right)\zeta_{M}\left(s+\frac{1}{2}\right)\;.
\end{equation}
By adding the terms obtained in (\ref{30}) through (\ref{34}) we obtain the result
\begin{equation}\label{39}
f(s,\beta,0)\sim\frac{\beta^{2s}}{2}\sum_{l=-D}^{\infty}\Gamma\left(-s-\frac{l}{2}\right)\beta^{l}A_{\frac{D+l}{2}}^{M}
\zeta_{R}(-2s-l)-\frac{1}{4}\Gamma(s)\zeta_{M}(s)+\frac{\sqrt{\pi}}{4\beta}\Gamma\left(s+\frac{1}{2}\right)\zeta_{M}\left(s+\frac{1}{2}\right)\;,
\end{equation}

Similarly to the previous case, when either $s=k$, $k\in\mathbb{Z}$, or $s=\pm (2n+1)/2$, $n\in\mathbb{N}_{0}$, the integrand in (\ref{11}) will have
simple as well as double poles. We start by considering the values $s=n$, with $n\in\mathbb{N}^{+}$, and $D=2d$. The simple and double poles of the integrand
in (\ref{11}) coincide with the ones listed just before (\ref{25b}). In addition to those poles, we have that the point $t=1/2$ is a simple pole when $n\geq d$ and
a double pole when $1\leq n\leq d-1$. By closing the contour of integration to the left, the integral in (\ref{11}) can be computed by using Cauchy's residue theorem
to obtain
\begin{eqnarray}\label{39a}
\lefteqn{f_{D=2d}(n,\beta,0)\sim\frac{1}{4}\left[-(n-1)!\textrm{FP}\,\zeta_{M}(n)+A_{d-n}^{M}(\gamma-\Psi(n)+2\ln\beta)\right]\Theta(d-n)}\nonumber\\
& &+\Theta(d-n-1)\Bigg[\frac{\beta^{2n-2d}}{2}\sum_{j=0}^{d-n-1}\Gamma(-n+d-j)\beta^{2j}A_{j}^{M}\zeta_{R}(-2n+2d-2j)\nonumber\\
&&+\frac{\sqrt{\pi}}{4\beta}\Gamma\left(n+\frac{1}{2}\right)\textrm{FP}\,\zeta_{M}\left(n+\frac{1}{2}\right)-\frac{\sqrt{\pi}}{2\beta}A_{d-n-\frac{1}{2}}^{M}\Bigg(\ln 4\beta-\sum_{k=1}^{n}\frac{1}{2k-1}\Bigg)\Bigg]\nonumber\\
&&+\frac{\beta^{2n+1}}{2}\sum_{m=-d \atop m\neq -n-1}^{\infty}\Gamma\left(-n-\frac{2m+1}{2}\right)\beta^{2m}A^{M}_{d+m+\frac{1}{2}}\zeta_{R}(-2n-2m-1)\nonumber\\
&&+\beta^{2n-2d}\sum_{j=\max\{d-n,0\}}^{d-1}\frac{(-1)^{n-d+j}}{(n-d+j)!}\beta^{2j}A^{M}_{j}\zeta_{R}^{\prime}(-2n+2d-2j)
+\sum_{m=n}^{\infty}\frac{(-1)^{m}}{m!}\beta^{2m}A^{M}_{d+m-n}\zeta_{R}^{\prime}(-2m)\nonumber\\
&&-\frac{1}{4}(n-1)!\zeta_{M}(n)\Theta(n-d-1)+\frac{\sqrt{\pi}}{4\beta}\Gamma\left(n+\frac{1}{2}\right)\zeta_{M}\left(n+\frac{1}{2}\right)\Theta(n-d)\;.
\end{eqnarray}
For $s=n$, $n\in\mathbb{N}^{+}$, and $D=2d+1$, the simple and double poles of the integrand in (\ref{11}) are the same as the ones given before the expression
(\ref{25c}). Moreover, $t=1/2$ is a simple pole for $n\geq d+1$ and a double pole for $1\leq n\leq d$. In this case we obtain
\begin{eqnarray}\label{39b}
\lefteqn{f_{D=2d+1}(n,\beta,0)\sim\frac{1}{4}\Bigg[-(n-1)!\textrm{FP}\,\zeta_{M}(n)+A_{d-n+\frac{1}{2}}^{M}(\gamma-\Psi(n)+2\ln\beta)}\nonumber\\
&&+\frac{\sqrt{\pi}}{\beta}\Gamma\left(n+\frac{1}{2}\right)\textrm{FP}\,\zeta_{M}\left(n+\frac{1}{2}\right)-\frac{2\sqrt{\pi}}{\beta}A_{d-n}^{M}\Bigg(\ln 4\beta-\sum_{k=1}^{n}\frac{1}{2k-1}\Bigg)\Bigg]\Theta(d-n)\nonumber\\
& &+\Theta(d-n-1)\frac{\beta^{2n-2d}}{2}\sum_{p=0}^{d-n-1}\Gamma(-n+d-p)\beta^{2p}A_{p+\frac{1}{2}}^{M}\zeta_{R}(-2n+2d-2p)\nonumber\\
&&+\frac{\beta^{2n+1}}{2}\sum_{m=-d-1 \atop m\neq -n-1}^{\infty}\Gamma\left(-n-\frac{2m+1}{2}\right)\beta^{2m}A^{M}_{d+m+1}\zeta_{R}(-2n-2m-1)\nonumber\\
&&+\beta^{2n-2d}\sum_{p=\max\{d-n,0\}}^{d-1}\frac{(-1)^{n-d+p}}{(n-d+p)!}\beta^{2p}A^{M}_{p+\frac{1}{2}}\zeta_{R}^{\prime}(-2n+2d-2p)
+\sum_{m=n}^{\infty}\frac{(-1)^{m}}{m!}\beta^{2m}A^{M}_{d+m-n+\frac{1}{2}}\zeta_{R}^{\prime}(-2m)\nonumber\\
&&-\frac{1}{4}\Theta(n-d-1)\left[(n-1)!\zeta_{M}(n)-\frac{\sqrt{\pi}}{\beta}\Gamma\left(n+\frac{1}{2}\right)\zeta_{M}\left(n+\frac{1}{2}\right)\right]\;.
\end{eqnarray}
When $s=-n$, with $n\in\mathbb{N}_{0}$, and $M$ is either an even or odd-dimensional manifold, the poles of the integrand of (\ref{11}) are
located at the points indicated before equation (\ref{25d}) with the addition of a double pole at $t=1/2$. In this case the integral (\ref{11}) gives
\begin{eqnarray}\label{39c}
\lefteqn{f(-n,\beta,0)\sim\frac{1}{4}A^{M}_{\frac{D}{2}+n}\left[\gamma-\Psi(1+n)+2\ln\beta\right]-\frac{1}{4}\frac{(-1)^{n}}{n!}\zeta_{M}^{\prime}(-n)}\nonumber\\
&&+\frac{\beta^{-2n-D}}{2}\sum_{k=0}^{D-2}\Gamma\left(n+\frac{D-k}{2}\right)\beta^{k}A^{M}_{\frac{k}{2}}\zeta_{R}(2n+D-k)
+\Theta(n-1)\frac{\beta^{-2n-1}}{2}\Gamma\left(n+\frac{1}{2}\right)\beta^{k}A^{M}_{\frac{D-1}{2}}\zeta_{R}(2n+1)\nonumber\\
&&+\frac{\beta^{-2n+1}}{2}\sum_{m=0 \atop m\neq n-1}^{\infty}\Gamma\left(n-\frac{2m+1}{2}\right)\beta^{2m}A^{M}_{\frac{D+2m+1}{2}}\zeta_{R}(2n-2m-1)
+\sum_{m=0}^{\infty}\frac{(-1)^{m}}{m!}\beta^{2m}A^{M}_{\frac{D}{2}+m+n}\zeta_{R}^{\prime}(-2m)\nonumber\\
&&+\frac{\Theta(n-1)}{2}\sum_{m=1}^{n}(m-1)!\beta^{-2m}A^{M}_{\frac{D}{2}+n-m}\zeta_{R}(2m)
+\frac{\sqrt{\pi}}{4\beta}\Gamma\left(\frac{1}{2}-n\right)\textrm{FP}\,\zeta_{M}\left(\frac{1}{2}-n\right)\nonumber\\
&&-\frac{\sqrt{\pi}}{2\beta}A^{M}_{\frac{D+2n-1}{2}}\Bigg[\ln 4\beta-\sum_{k=1}^{n}\frac{1}{2k-1}\Bigg]\;.
\end{eqnarray}

We next focus on the values $s=(2n+1)/2$, $n\in\mathbb{N}_{0}$, and $D=2d$. In this case, the integrand in (\ref{11}) has poles
at the points listed before (\ref{25e}). In addition, $t=1/2$ is a simple pole if $n\geq d$ and a double pole if $0\leq n\leq d-1$.
Taking into account the above poles, the integral in (\ref{11}) leads to the result
\begin{eqnarray}\label{39d}
\lefteqn{f_{D=2d}\left(\frac{2n+1}{2},\beta,0\right)\sim\frac{1}{4}\Bigg[-\Gamma\left(n+\frac{1}{2}\right)\textrm{FP}\,\zeta_{M}\left(n+\frac{1}{2}\right)+A_{d-n-\frac{1}{2}}^{M}\left(\gamma-\Psi\left(n+\frac{1}{2}\right)+2\ln\beta\right)}\nonumber\\
&&+\frac{\sqrt{\pi}}{\beta}n!\textrm{FP}\,\zeta_{M}(n+1)-\frac{2\sqrt{\pi}}{\beta}A^{M}_{d-n-1}\left(\ln 2\beta-2H_{n}\right)\Bigg]\Theta(d-n-1)\nonumber\\
&&+\Theta(d-n-2)\frac{\beta^{2n-2d+2}}{2}\sum_{j=0}^{d-n-2}\Gamma(-n+d-j-1)\beta^{2j}A_{j+\frac{1}{2}}^{M}\zeta_{R}(-2n+2d-2j-2)\nonumber\\
&&+\frac{\beta^{2n+1}}{2}\sum_{m=-d \atop m\neq -n-1}^{\infty}\Gamma\left(-n-\frac{2m+1}{2}\right)\beta^{2m}A^{M}_{d+m}\zeta_{R}(-2n-2m-1)\nonumber\\
&&+\beta^{2n-2d+2}\sum_{j=\max\{d-n-1,0\}}^{d-1}\frac{(-1)^{n-d+j+1}}{(n-d+j+1)!}\beta^{2j}A^{M}_{j+\frac{1}{2}}\zeta_{R}^{\prime}(-2n+2d-2j-2)\\
&&+\sum_{m=n+1}^{\infty}\frac{(-1)^{m}}{m!}\beta^{2m}A^{M}_{d+m-n-\frac{1}{2}}\zeta_{R}^{\prime}(-2m)
-\frac{1}{4}\Theta(n-d)\left[\Gamma\left(n+\frac{1}{2}\right)\zeta_{M}\left(n+\frac{1}{2}\right)-\frac{\sqrt{\pi}}{\beta}n!\zeta_{M}(n+1)\right]\nonumber ,
\end{eqnarray}
where $H_{n}$ denotes the $n$-th harmonic number. For $s=(2n+1)/2$, $n\in\mathbb{N}_{0}$, and $D=2d+1$,
the integrand in (\ref{11}) develops poles at the points indicated above (\ref{25f}).
Furthermore, the point $t=1/2$ becomes a simple pole when $n\geq d+1$ and a double pole when $0\leq n\leq d$. In this case the integral (\ref{11}) gives
\begin{eqnarray}\label{39e}
f_{D=2d+1}\left(\frac{2n+1}{2},\beta,0\right)&\sim&\frac{1}{4}\left[-\Gamma\left(n+\frac{1}{2}\right)\textrm{FP}\,\zeta_{M}\left(n+\frac{1}{2}\right)
+A_{d-n}^{M}\left(\gamma-\Psi\left(n+\frac{1}{2}\right)+2\ln\beta\right)\right]\Theta(d-n)\nonumber\\
&+&\Theta(d-n-1)\Bigg[\frac{\beta^{2n-2d}}{2}\sum_{j=0}^{d-n-1}\Gamma(-n+d-j)\beta^{2j}A_{j}^{M}\zeta_{R}(-2n+2d-2j)\nonumber\\
&+&\frac{\sqrt{\pi}}{4\beta}n!\textrm{FP}\,\zeta_{M}(n+1)-\frac{\sqrt{\pi}}{2\beta}A^{M}_{d-n-\frac{1}{2}}\left(\ln 2\beta-2H_{n}\right)\Bigg]\nonumber\\
&+&\frac{\beta^{2n+1}}{2}\sum_{m=-d\atop m\neq -n-1}^{\infty}\Gamma\left(-n-\frac{2m+1}{2}\right)\beta^{2m}A^{M}_{d+m+\frac{1}{2}}\zeta_{R}(-2n-2m-1)\nonumber\\
&+&\beta^{2n-2d}\sum_{j=\max\{d-n,0\}}^{d}\frac{(-1)^{n-d+j}}{(n-d+j)!}\beta^{2j}A^{M}_{j}\zeta_{R}^{\prime}(-2n+2d-2j)\nonumber\\
&+&\sum_{m=n+1}^{\infty}\frac{(-1)^{m}}{m!}\beta^{2m}A^{M}_{d+m-n}\zeta_{R}^{\prime}(-2m)-\frac{1}{4}\Gamma\left(n+\frac{1}{2}\right)\zeta_{M}\left(n+\frac{1}{2}\right)\Theta(n-d-1)\;,\nonumber\\
&+&\frac{\sqrt{\pi}}{4\beta}n!\zeta_{M}(n+1)\Theta(n-d)
\end{eqnarray}
Finally, when $s=-(2n+1)/2$, with $n\in\mathbb{N}_{0}$, and $M$ is either an even or odd-dimensional manifold, the integrand in (\ref{11}) has poles located at the
points indicated before equation (\ref{25g}) and a double pole at $t=1/2$. In this case the integral (\ref{11}) gives the result
\begin{eqnarray}\label{39f}
\lefteqn{f\left(-\frac{2n+1}{2},\beta,0\right)\sim-\frac{1}{4}A^{M}_{\frac{D+2n+1}{2}}\left[\Psi\left(-\frac{2n+1}{2}\right)-\gamma-2\ln\beta\right]
-\frac{1}{4}\Gamma\left(-\frac{2n+1}{2}\right)\textrm{FP}\,\zeta_{M}\left(-\frac{2n+1}{2}\right)}\nonumber\\
&&+\frac{\beta^{-2n-D-1}}{2}\sum_{k=0}^{D-1}\Gamma\left(n+\frac{D-k+1}{2}\right)\beta^{k}A^{M}_{\frac{k}{2}}\zeta_{R}(2n+D-k+1)\nonumber\\
&&+\frac{\beta^{-2n-1}}{2}\sum_{m=0\atop m\neq n}^{\infty}\Gamma\left(n-m+\frac{1}{2}\right)\beta^{2m}A^{M}_{\frac{D}{2}+m}\zeta_{R}(2n-2m+1)
+\sum_{m=0}^{\infty}\frac{(-1)^{m}}{m!}\beta^{2m}A^{M}_{\frac{D+1}{2}+m+n}\zeta_{R}^{\prime}(-2m)\nonumber\\
&&+\frac{\Theta(n-1)}{2}\sum_{m=1}^{n}(m-1)!\beta^{-2m}A^{M}_{\frac{D+1}{2}+n-m}\zeta_{R}(2m)+\frac{\sqrt{\pi}}{4\beta}\frac{(-1)^{n}}{n!}\zeta^{\prime}_{M}(-n)
+\frac{\sqrt{\pi}}{4\beta}A^{M}_{\frac{D}{2}+n}\left(-2\ln 2\beta+H_{n}\right)\;.\nonumber\\
\end{eqnarray}

\section{Small Parameter Expansion of the Bessel Series}\label{sec2a}

Let us now turn our attention to the series (\ref{2}) and (\ref{3}) defined in the Introduction. By utilizing the complex integral representation (\ref{6})
for the modified Bessel function of the second kind in the expression for $h(s,\beta,B)$ in (\ref{2}), we obtain
\begin{equation}\label{40}
h(s,\beta,B)=\frac{1}{8\pi i}\int_{c-i\infty}^{c+i\infty}\Gamma(t)\Gamma(t+s)\beta^{-2t}\left[\textrm{Li}_{2t}\left(e^{2\pi i B}\right)+\textrm{Li}_{2t}\left(e^{-2\pi i B}\right)\right]\diff t\;,
\end{equation}
with $c>\textrm{max}\{0,-\Re(s)\}$. The integrand can be shown to have only simple poles when $s\neq k$, with $k\in\mathbb{Z}$, positioned at the points
$t=-s-n$ and $t=-n$ where $n\in\mathbb{N}_{0}$. For $s=k$, instead, the integrand develops both simple and double poles.
By assuming that $s\neq k$ we close the integration contour to the left in (\ref{40}) and apply Cauchy's residue theorem to obtain the following
contribution
\begin{equation}\label{41}
h_{1}(s,\beta,B)=\frac{\beta^{2s}}{4}\sum_{n=0}^{\infty}\frac{(-1)^{n}}{n!}\Gamma(-s-n)\beta^{2n}
\left[\textrm{Li}_{-2s-2l}\left(e^{2\pi i B}\right)+\textrm{Li}_{-2s-2l}\left(e^{-2\pi i B}\right)\right]
\end{equation}
from the poles at $t=-s-n$. From the poles at $t=-n$ we have, instead,
\begin{equation}\label{42}
h_{2}(s,\beta,B)=-\frac{1}{4}\Gamma(s)\;.
\end{equation}
By adding the results in (\ref{41}) and (\ref{42}) we have
\begin{equation}\label{43}
h(s,\beta,B)\sim\frac{\beta^{2s}}{4}\sum_{n=0}^{\infty}\frac{(-1)^{n}}{n!}\Gamma(-s-n)\beta^{2n}
\left[\textrm{Li}_{-2s-2n}\left(e^{2\pi i B}\right)+\textrm{Li}_{-2s-2n}\left(e^{-2\pi i B}\right)\right]-\frac{1}{4}\Gamma(s)\;.
\end{equation}

When $s=n$, with $n\in\mathbb{N}^{+}$, the integrand in (\ref{40}) has poles at $t=-j$, $j\in\mathbb{N}_{0}$, which are simple
when $j=\{0,\ldots,n-1\}$ and double when $j\geq n$. In this case the integral (\ref{40}) gives
\begin{equation}\label{43a}
h(n,\beta,B)\sim\frac{(-1)^{n}}{2}\sum_{j=n}^{\infty}\frac{\beta^{2j}}{j!(j-n)!}\left[\textrm{Li}^{\prime}_{-2j}\left(e^{2\pi i B}\right)+\textrm{Li}^{\prime}_{-2j}\left(e^{-2\pi i B}\right)\right]-\frac{(n-1)!}{4}\;.
\end{equation}
For $s=-n$, with $n\in\mathbb{N}_{0}$, the integrand in (\ref{40}) has simple poles at $t=j$, with $j=\{1,\ldots,n\}$, and double poles for $t=-j$, with $j\geq 0$.
The integral (\ref{40}) then provides the result
\begin{eqnarray}\label{43b}
h(-n,\beta,B)&\sim&\frac{1}{2}\frac{(-1)^{n}}{n!}\left(\gamma+\ln\beta-2H_{n}\right)+\frac{(-1)^{n}}{2}\sum_{j=0}^{\infty}\frac{\beta^{2j}}{j!(j+n)!}\left[\textrm{Li}^{\prime}_{-2j}\left(e^{2\pi i B}\right)+\textrm{Li}^{\prime}_{-2j}\left(e^{-2\pi i B}\right)\right]\nonumber\\
&+&\frac{1}{4}\sum_{j=1}^{n}\frac{(-1)^{n-j}(j-1)!}{(n-j)!}\beta^{-2j}\left[\textrm{Li}_{2j}\left(e^{2\pi i B}\right)+\textrm{Li}_{2j}\left(e^{-2\pi i B}\right)\right]\;.
\end{eqnarray}

For $B=0$ the series $h(s,\beta,B)$ in (\ref{2}) becomes
\begin{equation}\label{48}
h(s,\beta,0)=\sum_{m=1}^{\infty}(m\beta)^{s}K_{-s}(2m\beta)\;.
\end{equation}
The integral representation (\ref{6}) allows us to rewrite (\ref{48}) as follows
\begin{equation}\label{49}
h(s,\beta,0)=\frac{1}{4\pi i}\int_{c-i\infty}^{c+i\infty}\Gamma(t)\Gamma(t+s)\beta^{-2t}\zeta_{R}(2t)\diff t\;,
\end{equation}
where $c>\textrm{max}\{1/2,-\Re(s)\}$. The integrand in (\ref{49}) develops poles at the points
$t=-s-n$, $t=-n$, and at $t=1/2$ with $n\in\mathbb{N}_{0}$. These poles are simple for values of $s\neq k$, with $k\in\mathbb{Z}$, and $s\neq -(2n+1)/2$, with
$n\in\mathbb{N}_{0}$. For all $s\in\mathbb{C}$ except for the integers and negative half-integers we close the contour of integration to the left and we obtain, from the first set of poles, the contribution
\begin{equation}\label{50}
h_{1}(s,\beta,0)=\frac{\beta^{2s}}{2}\sum_{n=0}^{\infty}\frac{(-1)^{n}}{n!}\Gamma(-s-n)\beta^{2n}
\zeta_{R}(-2s-2n)\;.
\end{equation}
From the poles at $t=-n$ we have that $h_{2}(s,\beta,0)=h_{2}(s,\beta,B)$, while the pole at $t=1/2$ leads to the result
\begin{equation}\label{51}
h_{3}(s,\beta,0)=\frac{\sqrt{\pi}}{4\beta}\Gamma\left(s+\frac{1}{2}\right)\;.
\end{equation}
By adding the contributions from all the poles we obtain the expansion
\begin{equation}\label{52}
h(s,\beta,0)\sim\frac{\beta^{2s}}{2}\sum_{n=0}^{\infty}\frac{(-1)^{n}}{n!}\Gamma(-s-n)\beta^{2n}
\zeta_{R}(-2s-2n)-\frac{1}{4}\Gamma(s)+\frac{\sqrt{\pi}}{4\beta}\Gamma\left(s+\frac{1}{2}\right)\;.
\end{equation}

Now, for $s=m$, $m\in\mathbb{N}^{+}$, the integrand in (\ref{49}) has simple poles for $t=-j$, negative integers, with $j=\{0,\ldots,n-1\}$ and at $t=1/2$.
Double poles are present, instead, for $t=-j$ where $j\geq n$. In this case the integral (\ref{49}) becomes
\begin{eqnarray}\label{52a}
h(n,\beta,0)\sim(-1)^{n}\sum_{j=n}^{\infty}\frac{\beta^{2j}}{j!(j-n)!}\zeta_{R}^{\prime}(-2j)-\frac{1}{4}\Gamma(n)+\frac{\sqrt{\pi}}{4\beta}\Gamma\left(n+\frac{1}{2}\right)\;.
\end{eqnarray}
For $s=-m$ with $m\in\mathbb{N}_{0}$, the integrand in (\ref{49}) has simple poles for $t=j$, $j=\{1,\ldots,n\}$ and $t=1/2$, and double poles for $t=-j$ with
$j\in\mathbb{N}_{0}$. The integral (\ref{49}) then gives
\begin{eqnarray}\label{52b}
h(-m,\beta,0)&\sim&\frac{1}{2}\frac{(-1)^{m}}{m!}\left(\gamma+\ln\beta-2H_{m}\right)+(-1)^{m}\sum_{j=0}^{\infty}\frac{\beta^{2j}}{j!(j+m)!}\zeta_{R}^{\prime}(-2j)
+\frac{\sqrt{\pi}}{4\beta}\Gamma\left(\frac{1}{2}-m\right)\nonumber\\
&+&\frac{1}{2}\sum_{j=1}^{m}\frac{(-1)^{m-j}(j-1)!}{(m-j)!}\beta^{-2j}\zeta_{R}(2j)\;.
\end{eqnarray}
Lastly, when $s=-(2n+1)/2$, the integrand in (\ref{49}) has simple poles for $t=-m$, with $m \in\mathbb{N}_{0}$, and $t=-m+n+1/2$ with $m\neq n$.
In addition, a double pole develops at $t=1/2$. In this case we have
\begin{eqnarray}\label{52c}
h\left(-\frac{2n+1}{2},\beta,0\right)&\sim&\frac{\beta^{-2n-1}}{2}\sum_{m=0\atop m\neq n}^{\infty}\frac{(-1)^{m}}{m!}\Gamma\left(-m+n+\frac{1}{2}\right)\beta^{2m}\zeta_{R}(-2m+2n+1)\nonumber\\
&-&\frac{1}{4}\Gamma\left(\frac{1}{2}-n\right)-\frac{\sqrt{\pi}}{4\beta}\frac{(-1)^{n}}{n!}\left(2\ln 2\beta-H_{n}\right)\;.
\end{eqnarray}

The last Bessel series considered in this work is $g(s,\beta)$ defined in (\ref{3}). By making use of the integral representation (\ref{6})
we write the series (\ref{3}) as
\begin{equation}\label{59}
g(s,\beta)=\frac{1}{4\pi i}\int_{c-i\infty}^{c+i\infty}\Gamma(t)\Gamma(t+s)\beta^{-2t}\zeta_{E}(s+t)\diff t\;,
\end{equation}
where $c>\textrm{max}\{0,d/2-\Re(s)\}$ and $\zeta_{E}(u)$ denotes the Epstein zeta function \cite{epst03-56-615,epst07-63-205}
\begin{equation}\label{69a}
\zeta_{E}(u)={\sum_{{\bf n}\in \mathbb{Z}^{d}}}^{\prime}\left(n_{1}^{2}+\cdots+n_{d}^{2}\right)^{-u}\;,
\end{equation}
which is valid for $\Re(u)>d/2$. The following reflection formula for the Epstein zeta function (\ref{69a})
\begin{equation}\label{69}
\zeta_{E}(u)=\frac{\pi^{2u-\frac{d}{2}}}{\Gamma(u)}\Gamma\left(\frac{d}{2}-u\right)\zeta_{E}\left(\frac{d}{2}-u\right)
\end{equation}
provides the analytic continuation of $\zeta_{E}(s)$ to the entire complex plane with a single simple pole at the point $u=d/2$ \cite{epst03-56-615,epst07-63-205}.
When $s\neq k$, $k\in\mathbb{Z}$, and $s\neq (2n+1)/2$, with the latter holding only for the case of odd $d$,
the integrand in (\ref{59}) possesses simple poles at the points $t=-s-n$, $t=-n$, and $t=d/2-s$ where $n\in\mathbb{N}_{0}$.
Under the above assumptions, by closing the integration contour to the left we obtain from the first set of poles
\begin{equation}\label{70}
g_{1}(s,\beta)=\frac{\beta^{2s}}{2}\sum_{n=0}^{\infty}\frac{(-1)^{n}}{n!}\Gamma(-s-n)\beta^{2n}\zeta_{E}(-n)\;.
\end{equation}
Since one can prove from (\ref{69}) that $\zeta_{E}(-p)=0$ for $p\in\mathbb{N}$ and $\zeta_{E}(0)=-1$, we can write (\ref{70}) as
\begin{equation}\label{71}
g_{1}(s,\beta)=-\frac{\beta^{2s}}{2}\Gamma(-s)\;.
\end{equation}
From the poles at $t=-n$ we obtain the contribution
\begin{equation}\label{72}
g_{2}(s,\beta)=\frac{1}{2}\sum_{n=0}^{\infty}\frac{(-1)^{n}}{n!}\Gamma(s-n)\beta^{2n}\zeta_{E}(s-n)\;,
\end{equation}
and from the ones at $t=d/2-s$ we have
\begin{equation}\label{73}
g_{3}(s,\beta)=\frac{\beta^{2s-d}\pi^{\frac{d}{2}}}{2}\Gamma\left(\frac{d}{2}-s\right)\;,
\end{equation}
which can be proved by noticing that
\begin{equation}
\textrm{Res}\,\zeta_{E}\left(\frac{d}{2}\right)=\frac{\pi^{\frac{d}{2}}}{\Gamma\left(\frac{d}{2}\right)}\;.
\end{equation}
Adding the above results allows us to obtain
\begin{equation}\label{74}
g(s,\beta)\sim\frac{1}{2}\sum_{n=0}^{\infty}\frac{(-1)^{n}}{n!}\Gamma(s-n)\beta^{2n}\zeta_{E}(s-n)-\frac{\beta^{2s}}{2}\Gamma(-s)
+\frac{\beta^{2s-d}\pi^{\frac{d}{2}}}{2}\Gamma\left(\frac{d}{2}-s\right)\; .
\end{equation}

To compute the small-$\beta$ expansion of $g(s,\beta)$ when $s$ is an integer or a positive half-integer, it is convenient
to distinguish between even and odd values of $d$.

We consider, first, the case of even $d$, namely $d=2l$ with $l\in\mathbb{N}^{+}$.
For $s=n$, $n\in\mathbb{N}_{0}$, and  the integrand in (\ref{59}) has simple poles for $t=-j$, $j=\{0,\ldots,n-1\}$ with $j\neq n-l$, and double
poles at $t=-j$, with $j\geq n$. In addition, $t=l-n$ is a simple pole when $l\geq n+1$, while it becomes a double pole when $l=\{1,\ldots,n\}$.
In this case the integral (\ref{59}) gives
\begin{eqnarray}\label{74a}
g_{d=2l}(n,\beta)&\sim&\frac{1}{2}\sum_{j=0\atop j\neq n-l}^{n-1}\frac{(-1)^{j}}{j!}\Gamma(n-j)\beta^{2j}\zeta_{E}(n-j)
+\frac{(-1)^{n}}{2}\sum_{j=n}^{\infty}\frac{\beta^{2j}}{j!(j-n)!}\zeta_{E}^{\prime}(n-j)\nonumber\\
&+&\frac{(-1)^{n}}{n!}\beta^{2n}\left(\gamma+\ln\beta-2H_{n}\right)+\Theta(l-n-1)\frac{\beta^{2n-2l}\pi^{l}}{2}\Gamma(l-n)\nonumber\\
&+&\Theta(n-l)\frac{(-1)^{n-l}\beta^{2n-2l}\pi^{l}}{2(n-l)!}\left[\pi^{-l}(l-1)!\textrm{FP}\,\zeta_{E}(l)+\Psi(n-l+1)+\Psi(l)-2\ln\beta\right]\;.
\end{eqnarray}
For $s=-n$, with $n\in\mathbb{N}^{+}$, the integrand in (\ref{59}) has simple poles at $t=j$, with $j=\{1,\ldots,n\}$ and at $t=l+n$.
Double poles, instead, are present at the points $t=-j$, $j\in\mathbb{N}_{0}$. The integral (\ref{59}) then becomes
\begin{eqnarray}\label{74b}
g_{d=2l}(-n,\beta)\sim-\frac{(n-1)!}{2\beta^{2n}}+\frac{\pi^{l}}{2\beta^{2l+2n}}\Gamma(n+l)+\frac{(-1)^{n}}{2}\sum_{j=0}^{\infty}\frac{\beta^{2j}}{j!(n+j)!}\zeta_{E}^{\prime}(-j-n)\;.
\end{eqnarray}
We now analyze the case $d=2l+1$, with $l\in\mathbb{N}_{0}$. For $s=n$, $n\in\mathbb{N}_{0}$, the integrand in (\ref{59}) has simple poles
at $t=-j$, with $j=\{0,\ldots,n-1\}$, and at $t=l-n+1/2$. Double poles appear at the points $t=-j$, with $j\geq n$. In this case we obtain
\begin{eqnarray}\label{74c}
g_{d=2l+1}(n,\beta)&\sim&\frac{1}{2}\sum_{j=0}^{n-1}\frac{(-1)^{j}}{j!}\Gamma(n-j)\beta^{2j}\zeta_{E}(n-j)
+\frac{(-1)^{n}}{2}\sum_{j=n}^{\infty}\frac{\beta^{2j}}{j!(j-n)!}\zeta_{E}^{\prime}(n-j)\nonumber\\
&+&\frac{(-1)^{n}}{n!}\beta^{2n}\left(\gamma+\ln\beta-2H_{n}\right)+\frac{\pi^{l+\frac{1}{2}}}{2\beta^{2l-2n+1}}\Gamma\left(l-n+\frac{1}{2}\right)\;.
\end{eqnarray}
When $s=-n$, with $n\in\mathbb{N}^{+}$, the integrand in (\ref{59}) develops simple poles at $t=j$, with $j=\{1,\ldots,n\}$, and at $t=l+n+1/2$, and double
poles at $t=-j$, where $j\in\mathbb{N}_{0}$. The integral (\ref{59}) can then be computed to give
\begin{eqnarray}\label{74d}
g_{d=2l+1}(-n,\beta)\sim-\frac{(n-1)!}{2\beta^{2n}}+\frac{\pi^{l+\frac{1}{2}}}{2\beta^{2l+2n+1}}\Gamma\left(n+l+\frac{1}{2}\right)+\frac{(-1)^{n}}{2}\sum_{j=0}^{\infty}\frac{\beta^{2j}}{j!(n+j)!}\zeta_{E}^{\prime}(-j-n)\;.
\end{eqnarray}
For $s=(2n+1)/2$ with $n\in\mathbb{N}_{0}$, the integrand in (\ref{59}) has simple poles at $t=-j$, $j\in\mathbb{N}_{0}$ with $j\neq n-l$,
and at $t=-m-n-1/2$, for $m\in\mathbb{N}_{0}$. Moreover, $t=-n+l$ is a simple pole for $l\geq n+1$, while  it is a double pole for $l=\{0,\ldots,n\}$.
The integral in (\ref{59}) then gives
\begin{eqnarray}\label{74e}
\lefteqn{g_{d=2l+1}\left(\frac{2n+1}{2},\beta\right)\sim\frac{1}{2}\sum_{j=0\atop j\neq n-l}^{\infty}\frac{(-1)^{j}}{j!}\Gamma\left(n-j+\frac{1}{2}\right)\beta^{2j}\zeta_{E}\left(n-j+\frac{1}{2}\right)-\frac{\beta^{2n+1}}{2}\Gamma\left(-n-\frac{1}{2}\right)}\nonumber\\
&&+\Theta(n-l)\frac{(-1)^{n-l}\beta^{2n-2l}\pi^{l+\frac{1}{2}}}{2(n-l)!}\left[\pi^{-l-\frac{1}{2}}\Gamma\left(l+\frac{1}{2}\right)\textrm{FP}\,\zeta_{E}\left(l+\frac{1}{2}\right)+\Psi(n-l+1)+\Psi\left(l+\frac{1}{2}\right)-2\ln\beta\right]\nonumber\\
&&+\Theta(l-n-1)\frac{\beta^{2n-2l}\pi^{l+\frac{1}{2}}}{2}\Gamma(l-n)\;.
\end{eqnarray}

\section{Specific Examples}\label{sec3}

The general results for the small-$\beta$ expansion of Bessel series of the form (\ref{1})-(\ref{3}) found in the previous
sections will now be used to study cases that are of particular interest in applications. The double Bessel series (\ref{1}), for instance,
appears in the expression of the spectral zeta function associated with the Laplace operator $L$ defined on quite general product manifolds.
In fact, let $M$ be a $D$-dimensional product manifold of the type $M=I\times U$, where $I=[a,b]\subset\mathbb{R}$ and $U=\mathbb{R}^{d}\times N$
with $N$ being a smooth Riemannian manifold with or without boundary for which we have $\textrm{dim}\,N=D-d-1$. In the appropriate coordinate system
the eigenvalue equation $L\varphi=\lambda^{2}\varphi$ can be separated and one can prove that the spectrum has the general form
\begin{equation}\label{83}
\lambda^{2}=\alpha^{2}+\gamma^{2}+\sum_{i=1}^{d}k_{i}^{2}\;,
\end{equation}
where $\alpha^{2}$ are the eigenvalues of the Laplacian on $N$, $\gamma^{2}$ are the ones associated with the Laplacian on $I$ and $k^{2}_{i}$
represents the continuous spectrum resulting from the Laplacian on $\mathbb{R}^{d}$. In this work we will restrict
our attention to the class of boundary conditions at the endpoints of $I$ that produce eigenvalues of the form
\begin{equation}\label{84}
\gamma^{2}=\frac{\pi^{2}}{\beta^{2}}(m+B)^{2}\;,
\end{equation}
with $\beta>0$ and $B\neq -m$. Depending on the specific boundary conditions chosen we can have either $m\in\mathbb{Z}$ or $m\in\mathbb{N}_{0}$.
We will continue our discussion under the assumption that $m\in\mathbb{Z}$. This is not restrictive since the spectral zeta function associated with a problem
for which $m\in\mathbb{N}_{0}$ can be obtained from the one with $m\in\mathbb{Z}$.
The eigenvalues (\ref{83}) with the definition (\ref{84}) define the following spectral zeta function density
\begin{equation}\label{85}
\zeta(s)=\frac{1}{(2\pi)^{d}}\int_{\mathbb{R}^{d}}\sum_{\alpha}\sum_{m\in\mathbb{Z}}\left(\alpha^{2}+\beta^{-2}\pi^{2}(m+B)^{2}+\sum_{i=1}^{d}k^{2}_{i}\right)^{-s}\diff {\bf k}\;.
\end{equation}
By using the integral \cite{gradshtein07}
\begin{equation}\label{86}
\int_{0}^{\infty}\left[1+\sum_{i=1}^{d}(rk_{i})^{2}\right]^{-s}\diff {\bf k}=2^{-d}\pi^{\frac{d}{2}}\frac{\Gamma\left(s-\frac{d}{2}\right)}{\Gamma(s)}r^{-d}\;,
\end{equation}
with $r>0$, we can perform the integral over the variables $k$ in (\ref{85}) and obtain
\begin{equation}\label{87}
\zeta(s)=\frac{\Gamma\left(s-\frac{d}{2}\right)}{(2\sqrt{\pi})^{d}\Gamma(s)}\sum_{\alpha}\sum_{m\in\mathbb{Z}}\left(\alpha^{2}+\beta^{-2}\pi^{2}(m+B)^{2}\right)^{-s+\frac{d}{2}}\;,
\end{equation}
which is valid for $\Re(s)>D/2$. To perform the analytic continuation of (\ref{87}) to values of $s$ to the left of the abscissa of convergence $\Re(s)=1/2$, we utilize a
method based on the Poisson summation formula.
For the argument of the double-series in (\ref{87}) we use the integral representation of the Gamma function to obtain
\begin{equation}\label{88}
\zeta(s)=\frac{1}{(2\sqrt{\pi})^{d}\Gamma(s)}\sum_{\alpha}\sum_{m\in\mathbb{Z}}
\int_{0}^{\infty}t^{s-\frac{d}{2}-1}\exp\left\{-t\left(\alpha^{2}+\beta^{-2}\pi^{2}(m+B)^{2}\right)\right\}\diff t\;.
\end{equation}
For $\Re(s)>D/2$, we first interchange the sums and the integral and then perform the sum over $m$ by using the Poisson resummation formula \cite{hille63}
\begin{equation}\label{89}
\sum_{m\in\mathbb{Z}}e^{-t\beta^{-2}\pi^{2}(m+B)^{2}}=\frac{\beta}{\sqrt{\pi t}}\left[1+2\sum_{m=1}^{\infty}\cos(2\pi mB)e^{-\frac{\beta^{2}m^{2}}{t}}\right]\;,
\end{equation}
to finally arrive at the expression
\begin{eqnarray}\label{90}
\zeta(s)&=&\frac{\beta}{2^{d}\pi^{\frac{d+1}{2}}\Gamma(s)}\int_{0}^{\infty}t^{s-\frac{d+1}{2}-1}\sum_{\alpha}e^{-t\alpha^{2}}\diff t\nonumber\\
&+&\frac{2\beta}{2^{d}\pi^{\frac{d+1}{2}}\Gamma(s)}\sum_{\alpha}\sum_{m=1}^{\infty}\cos(2\pi mB)\int_{0}^{\infty}t^{s-\frac{d+1}{2}-1}e^{-t\alpha^{2}-\frac{\beta^{2}m^{2}}{t}}\diff t\;.
\end{eqnarray}
The first term in (\ref{90}) is the Mellin transform of the trace of the heat kernel associated with the Laplacian on $N$ and, hence, is proportional to the spectral zeta function
\begin{equation}
\zeta_{N}(s)=\sum_{\alpha}\alpha^{-2s}\;,
\end{equation}
while the second term can be written in terms of modified Bessel functions of the second kind by virtue of the following integral representation \cite{gradshtein07}
\begin{equation}
K_{\nu}(xz)=\frac{z^{\nu}}{2}\int_{0}^{\infty}t^{-\nu-1}e^{-\frac{x}{2}\left(t+\frac{z^2}{t}\right)}\diff t\;.
\end{equation}

The last remarks allow us to write the expression
\begin{eqnarray}\label{91}
\zeta(s)&=&\frac{\beta}{2^{d}\pi^{\frac{d+1}{2}}}\frac{\Gamma\left(s-\frac{d+1}{2}\right)}{\Gamma(s)}\zeta_{N}\left(s-\frac{d+1}{2}\right)\nonumber\\
&+&\frac{\beta}{2^{d-2}\pi^{\frac{d+1}{2}}\Gamma(s)}\sum_{\alpha}\sum_{m=1}^{\infty}\cos(2\pi mB)\left(\frac{m\beta}{\alpha}\right)^{s-\frac{d+1}{2}}K_{-s+\frac{d+1}{2}}(2\alpha\beta m)\;,
\end{eqnarray}
which represents the analytic continuation of $\zeta(s)$ to a meromorphic function in the entire complex plane. Due to the exponential decay of the modified Bessel
function of the second kind, the meromorphic structure of $\zeta(s)$ is entirely encoded in the first term of (\ref{91}).
Spectral zeta functions of the form (\ref{91}) appear frequently in the literature, especially in the ambit of Kaluza-Klein theories, Casimir energy,
and in the study of finite temperature effects in quantum field theory \cite{ambj83,dowker84,elizalde94,kirsten09}.
By utilizing the result obtained in (\ref{25}) we find the following expression for the small-$\beta$ expansion of the spectral zeta function in (\ref{91})
\begin{equation}\label{92}
\zeta(s)\sim\frac{\beta^{2s-d}}{2^{d}\pi^{\frac{d+1}{2}}\Gamma(s)}\sum_{l=-D+d+1}^{\infty}\Gamma\left(-s+\frac{d+1-l}{2}\right)\beta^{l}A_{\frac{D-d-1+l}{2}}^{N}
\left[\textrm{Li}_{-2s+d+1-l}\left(e^{2\pi i B}\right)+\textrm{Li}_{-2s+d+1-l}\left(e^{-2\pi i B}\right)\right]\;,
\end{equation}
valid for all values of $s$. The expression (\ref{92}) can be used to study the small-$\beta$ behavior of quantities of interest. For instance,
the derivative at $s=0$ of (\ref{92}) with $\mu=iB$ would provide the high temperature expansion of the finite temperature one-loop effective action associated with a
quantum field propagating on $U$ with chemical potential $\mu$ (see e.g. \cite{elizalde94}).

As a further example we consider the following spectral zeta function
\begin{equation}\label{93}
\zeta(s,\beta)=-\frac{1}{2\pi^{\frac{D-1}{2}}}\frac{\Gamma\left(s-\frac{D-1}{2}\right)}{\Gamma(s)}\sum_{\alpha}\sum_{n=0}^{\infty}\left[\alpha^{2}+\frac{\pi^{2}}{\beta^{2}}\left(n+\frac{1}{2}\right)^{2}\right]^{-s+\frac{D-1}{2}}\;,
\end{equation}
which appears when studying the Casimir energy of fermions in the setting of a higher dimensional piston geometry of the type
$M^{D}\times N$ with $M^{D}$ a $D$-dimensional Euclidean space and $N$ a compact Riemannian manifold \cite{oiko14}.
Here, $\beta$ represents the length of the first chamber of the piston and $\alpha$ the eigenvalues of the Laplacian on $N$. We will also assume that $\textrm{dim}\,N=Q$.
By noticing that
\begin{equation}
\sum_{\alpha}\sum_{n=0}^{\infty}\left[\alpha^{2}+\frac{\pi^{2}}{\beta^{2}}\left(n+\frac{1}{2}\right)^{2}\right]^{-s+\frac{D-1}{2}}=
\frac{1}{2}\sum_{\alpha}\sum_{n\in\mathbb{Z}}\left[\alpha^{2}+\frac{\pi^{2}}{\beta^{2}}\left(n+\frac{1}{2}\right)^{2}\right]^{-s+\frac{D-1}{2}}\;,
\end{equation}
we have
\begin{equation}
\zeta(s,\beta)=-2^{D-3}\zeta(s)\;,
\end{equation}
with $\zeta(s)$ given in (\ref{87}) with the substitution $d=D-1$ and $B=1/2$. The last relation implies that
we can use the method outlined in this section to perform the analytic continuation of (\ref{93})  to obtain
\begin{eqnarray}\label{94}
\zeta(s,\beta)&=&-\frac{\beta}{4\pi^{\frac{D}{2}}}\frac{\Gamma\left(s-\frac{D}{2}\right)}{\Gamma(s)}\zeta_{N}\left(s-\frac{D}{2}\right)-\frac{\beta}{\pi^{\frac{D}{2}}\Gamma(s)}\sum_{\alpha}\sum_{n=1}^{\infty}(-1)^{n}\left(\frac{n\beta}{\alpha}\right)^{s-\frac{D}{2}}K_{-s+\frac{D}{2}}(2\alpha\beta n)\;.
\end{eqnarray}
By using the result (\ref{92}) with $B=1/2$ and the fact that
\begin{equation}
\textrm{Li}_{s}\left(e^{i\pi }\right)+\textrm{Li}_{s}\left(e^{-i\pi}\right)=2\left(2^{1-s}-1\right)\zeta_{R}(s)\;,
\end{equation}
one finds the following expansion of $\zeta(s,\beta)$ valid for $s\in\mathbb{C}$ when the size $\beta$ of the first chamber of the piston is small
\begin{equation}\label{95}
\zeta(s,\beta)\sim-\frac{\beta^{2s-D+1}}{2\pi^{\frac{D}{2}}\Gamma(s)}\sum_{l=-Q}^{\infty}\Gamma\left(-s+\frac{D-l}{2}\right)\beta^{l}A_{\frac{Q+l}{2}}^{N}
\left(2^{2s-D+l+1}-1\right)\zeta_{R}(-2s+D-l)\;.
\end{equation}
By setting $s=\varepsilon-1/2$ in (\ref{95}) we obtain an expression for the small-$\beta$ expansion of the Casimir energy
associated with the first chamber of the piston
\begin{equation}\label{96}
E_{C}\left(\varepsilon-\frac{1}{2},\beta\right)\sim\frac{\beta}{16\pi^{D+1}\varepsilon}A^{N}_{\frac{Q+D+1}{2}}+\frac{1}{8\pi^{\frac{D+1}{2}}}\sum_{l=-Q\atop{l\neq D+1}}^{\infty}
\Gamma\left(\frac{D-l+1}{2}\right)\beta^{l-D}A^{N}_{\frac{Q+l}{2}}\left(2^{l-D}-1\right)\zeta_{R}(D-l+1)\;,
\end{equation}
which, as $\epsilon \to 0$, is divergent as expected. By adding the Casimir energy contribution coming from the second chamber,
namely the expression (\ref{96}) with the replacemet $\beta\to L-\beta$, L being the total length of the piston, and by then differentiating
the resulting expression with respect to $\beta$ \cite{bord09b} we obtain the following Casimir force on the piston
\begin{equation}\label{97}
F_{C}(\beta)\sim\frac{1}{8\pi^{\frac{D+1}{2}}}\sum_{l=-Q\atop{l\neq D+1}}^{\infty}(l-D)\Gamma\left(\frac{D-l+1}{2}\right)\beta^{l-D}A^{N}_{\frac{Q+l}{2}}\left[(L-\beta)^{l-D-1}-\beta^{l-D-1}\right]\left(2^{l-D}-1\right)\zeta_{R}(D-l+1)\;,
\end{equation}
which is the correct expansion valid when the size $\beta$ of the first chamber is small.

We would like to consider, as an additional example, the following Bessel series
\begin{equation}\label{98}
S(m)=\sum_{n=1}^{\infty}\left(\frac{m}{nL}\right)^{\frac{D}{2}-1}K_{\frac{D}{2}-1}(nLm)\;,
\end{equation}
which appears in the expression for the one-loop renormalized mass in the Euclidean $\lambda\phi^{4}$ model compactified along one direction \cite{abreu05,mal03}.
Here, $L$ denotes the size of the one-dimensional compactified subspace of a $D$-dimensional Euclidean space, $m$ represents the physical mass of the field.
In the study of the critical behavior of this theory one is confronted with the task of analyzing the small-$m$ expansion of the series (\ref{98}) \cite{abreu05,mal03}.
By setting $m=2\beta/L$ we can rewrite (\ref{98}) in terms of the sum (\ref{48}) as follows
\begin{equation}\label{99}
S\left(\frac{2\beta}{L}\right)=\left(\frac{2}{L^{2}}\right)^{D/2-1}\beta^{D-2}\,h(s,\beta,0)\Big|_{s=1-\frac{D}{2}}\;,
\end{equation}
and, hence, we can use the small-$\beta$ expansion of $h(s,\beta,0)$ found in (\ref{52}). When the dimension of the Euclidean space is even, $D=2d$, $d\geq 1$,
we have, from (\ref{52b}) the expansion
\begin{eqnarray}\label{100}
h\left(1-d,\beta,0\right)&\sim&\frac{1}{2}\frac{(-1)^{d-1}}{(d-1)!}\left(\gamma+\ln\beta-2H_{d-1}\right)+(-1)^{d-1}\sum_{j=0}^{\infty}\frac{\beta^{2j}}{j!(j+d-1)!}\zeta_{R}^{\prime}(-2j)
+\frac{\sqrt{\pi}}{4\beta}\Gamma\left(\frac{3}{2}-d\right)\nonumber\\
&+&\frac{1}{2}\sum_{j=1}^{d-1}\frac{(-1)^{d-j-1}(j-1)!}{(d-j-1)!}\beta^{-2j}\zeta_{R}(2j)\;.
\end{eqnarray}
By applying this expansion to (\ref{99})
and by substituting back the variable $m$ we obtain for (\ref{98}) the expansion
\begin{eqnarray}\label{101}
S(m)&\sim&\frac{m^{2d-2}(-1)^{d-1}}{2^{d}(d-1)!}\left(\gamma+\ln\left(\frac{mL}{2}\right)-2H_{d-1}\right)+\frac{(-1)^{d-1}m^{2d-2}}{2^{d-1}}\sum_{j=0}^{\infty}\frac{(mL)^{2j}}{4^{j}j!(j+d-1)!}\zeta_{R}^{\prime}(-2j)\nonumber\\
&+&\frac{\sqrt{\pi}m^{2d-3}}{2^{d}L}\Gamma\left(\frac{3}{2}-d\right)+\frac{(-1)^{d-1}m^{2d-2}}{2^{d}}\sum_{j=1}^{d-1}\frac{4^{j}(-1)^{j}(j-1)!}{(d-j-1)!}(mL)^{-2j}\zeta_{R}(2j)\;.
\end{eqnarray}
valid when $D=2d$.

Let us next consider the case of odd-dimensional Euclidean space namely $D=2d+1$ with $d\geq 1$. In this case we use (\ref{52c}) to obtain
\begin{eqnarray}\label{102}
h\left(\frac{1}{2}-d,\beta,0\right)&\sim&\frac{\beta^{-2d+1}}{2}\sum_{m=0\atop m\neq d-1}^{\infty}\frac{(-1)^{m}}{m!}\Gamma\left(-m+d-\frac{1}{2}\right)\beta^{2m}\zeta_{R}(-2m+2d-1)\nonumber\\
&-&\frac{1}{4}\Gamma\left(\frac{3}{2}-d\right)-\frac{\sqrt{\pi}}{4\beta}\frac{(-1)^{d-1}}{(d-1)!}\left(2\ln 2\beta-H_{d-1}\right)\;.
\end{eqnarray}
This expansion can be used in (\ref{99}) to obtain, when substituting $\beta=mL/2$, the following small-$m$ expansion of (\ref{98})
valid for an odd-dimensional Euclidean space
\begin{eqnarray}\label{103}
S(m)&\sim&\frac{2^{d-\frac{1}{2}}}{L^{2d-1}}\sum_{m=0\atop m\neq d-1}^{\infty}\frac{(-1)^{m}}{4^{m}m!}\Gamma\left(-m+d-\frac{1}{2}\right)(mL)^{2m}\zeta_{R}(-2m+2d-1)\nonumber\\
&-&\frac{m^{2d-1}}{2^{d-\frac{1}{2}}}\Gamma\left(\frac{3}{2}-d\right)-\frac{\sqrt{\pi}m^{2d-2}}{2^{d+\frac{1}{2}}L}\frac{(-1)^{d-1}}{(d-1)!}\left(2\ln(mL)-H_{d-1}\right)\;.
\end{eqnarray}
The small-$m$ expansions of $S(m)$ in (\ref{98}) obtained in (\ref{101}) and (\ref{103}) are valid for any dimension $D$.
The expansion obtained in \cite{abreu05}, on the other hand, needs to be somehow regularized in the case $D=3$ which makes our results more suitable for the analysis
of $S(m)$ for any given dimension $D$.

\section{Concluding Remarks}\label{sec4}

In this work we have analyzed the asymptotic expansion of Bessel series valid when a given parameter,
which we have denoted by $\beta$, is small. Explicit and very general asymptotic expansions have been
obtained for both Bessel and double-Bessel series. The method used to obtain such general expansions
is based on the representation of the modified Bessel function of the second kind in terms of a
contour integral. After closing the contour of integration to the left, the integral is computed by using
Cauchy's residue theorem.
This method allowed us to find in a fairly straightforward way the desired small-$\beta$ expansion of the
Bessel series under consideration valid for all values of $s$.

In this paper we have focused our attention to the small-$\beta$ expansion of Bessel series of the form (\ref{1}),
(\ref{2}), and (\ref{3}). It is important to mention, however, that the method outlined in the previous sections can
easily be adapted to other types of Bessel series. For instance, if in the series (\ref{3}) we replace $|{\bf n}|$
with ${\bf n}\cdot {\bf r}$, with ${\bf r}\in\mathbb{R}^{d}_{+}$, and then sum over the lattice ${\bf n}\in\mathbb{N}^{d}_{+}$,
we obtain a new Bessel series whose small-$\beta$ expansion can be found by following the method described
in Section \ref{sec2} for $g(s,\beta)$. In this situation, however, the relevant zeta function appearing in the final
small-$\beta$ expansion would be the Barnes zeta function instead of the Epstein zeta function that is found in (\ref{74}).

The type of Bessel series considered in the previous sections appear frequently in the literature and
their expansions with respect to a small parameter prove to be of fundamental importance for obtaining
very useful information about special limiting cases. Unfortunately, however, such expansions are not
always performed correctly since the validity of the methods used is often dubious. The purpose of this
paper is then to serve as a guide for properly performing the small parameter expansion of infinite series
(including double series) containing modified Bessel functions of the second kind.

\begin{acknowledgments}
The research of G.F. is partially funded by the ORAU Ralph E. Powe Junior Faculty Enhancement Award.
\end{acknowledgments}

\end{document}